\documentstyle[12pt,twoside,epsf]{article}
\newcommand{\beq}{\begin{equation}}
\newcommand{\eeq}{\end{equation}}
\newcommand{\beqa}{\begin{eqnarray}}
\newcommand{\eeqa}{\end{eqnarray}}
\newcommand{\slsh}[1]{\, {\not {\! #1}}}
\newcommand{\slsha}[1]{\; {\not {\!\!\!\: #1}}}

\begin{document} 
\title{}
\author{}
\date{}
\maketitle
%\ \\[15mm]
%
%
% Ueberschrift und Autoren:
% =========================
%
{\LARGE \bf \raggedright Trace evaluation of matrix
determinants and inversion of 4 $\times$ 4 matrices in terms of Dirac 
covariants} \\
       
{\large \raggedright F.\ Kleefeld and M.\ Dillig} \\
{\small Institute for Theoretical Physics III, 
University of Erlangen--N\"urnberg, } \\
{\small Staudtstr.\ 7, D-91051 Erlangen, Germany} \\
{\small e--mail: kleefeld@theorie3.physik.uni--erlangen.de /
mdillig@theorie3.physik.uni--erlangen.de}\footnote{
Preprint-No.\ FAU-TP3-97/4} \\[5mm]

%
%
% Abstract:
% =========
%
\begin{figure}[hp]
\parbox{3.5cm}{
\epsfxsize=  3.5cm
\epsfysize=  4.5cm
\centerline{\epsffile{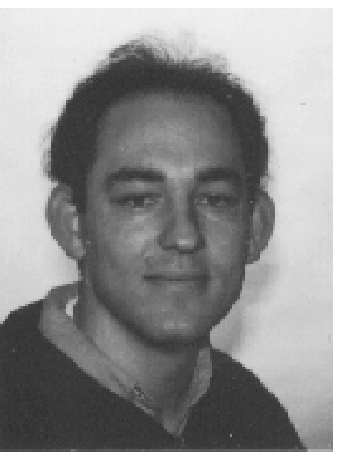}}}
\hfill
\parbox{12.0cm}{\noindent {\bf Abstract.} In the following short paper we list
some useful results concerning determinants and inverses of matrices. 
First we show, how to calculate determinants of $d \times d$ matrices, 
if their traces are known. As a next step
$4 \times 4$ matrices are expressed in terms of Dirac covariants. The third step
is the calculation of the corresponding inverse matrices in terms of Dirac
covariants.} 
\end{figure}

%
%
% Haupttext:
% ==========
%
%
% ==========================================================================
%
\section{Calculation of matrix determinants and permanents in terms of their
traces}

We denote the determinant of a quadratic $d \times d$ matrix A by:

\beq \det A = 
\left| \left| \begin{array}{ccc} A_{11} & \cdots & A_{1d} \\
                          \vdots &        & \vdots \\
                          A_{d1} & \cdots & A_{dd}
\end{array} \right| \right|
\eeq

The calculation of the determinant of $A$ for different dimensions $d$ is straight forward:

\beqa d=1 & \Rightarrow \quad \det A & = \quad A_{11} \nonumber \\
      d=2 & \Rightarrow \quad \det A & = \quad A_{11} A_{22} - A_{12} A_{21} \nonumber \\
      d=3 & \Rightarrow \quad \det A & = \quad A_{11} A_{22} A_{33}
                                             + A_{12} A_{23} A_{31} 
                                             + A_{13} A_{21} A_{32} - \nonumber \\
                                     & & \quad \; - A_{31} A_{22} A_{13}
                                               - A_{32} A_{23} A_{11} 
                                               - A_{33} A_{21} A_{12} \nonumber \\
 \cdots & & 
\eeqa

In a more unfamiliar way the determinants can be given by:

\beqa d=1 & \Rightarrow & \det A = \frac{1}{1!} \sum\limits_{i=1}^{1} 
\left| \left| \begin{array}{c} A_{ii} 
\end{array} \right| \right|
= \frac{1}{1!} \sum\limits_{i=1}^{1} A_{ii} \nonumber \\
 & & \nonumber \\
 d=2 & \Rightarrow & \det A = \frac{1}{2!} \sum\limits_{i,j=1}^{2} 
\left| \left| \begin{array}{cc} A_{ii} & A_{ij} \\
                                A_{ji} & A_{jj}
\end{array} \right| \right|
= \frac{1}{2!} \sum\limits_{i,j=1}^{2} (A_{ii} A_{jj} - A_{ij} A_{ji}) \nonumber \\ 
 & & \nonumber \\
 d=3 & \Rightarrow & \det A = \frac{1}{3!} \sum\limits_{i,j,k=1}^{3} 
\left| \left| \begin{array}{ccc} A_{ii} & A_{ij} & A_{ik} \\
                                 A_{ji} & A_{jj} & A_{jk} \\
                                 A_{ki} & A_{kj} & A_{kk} 
\end{array} \right| \right| = \nonumber \\
 & & \nonumber \\
 & & \qquad \;\; = \frac{1}{3!} \sum\limits_{i,j,k=1}^{3} \left(
                                                       A_{ii} A_{jj} A_{kk}
                                                     + A_{ij} A_{jk} A_{ki} 
                                                     + A_{ik} A_{ji} A_{kj} - \right. \nonumber \\                                          
                            & & \qquad \qquad \qquad \;\; \left. - A_{ki} A_{jj} A_{ik}
                                                     - A_{kj} A_{jk} A_{ii} 
                                                     - A_{kk} A_{ji} A_{ij} \right) \nonumber \\
 & & \nonumber \\
 d=4 & \Rightarrow & \det A = \frac{1}{4!} \sum\limits_{i,j,k,\ell=1}^{4} 
\left| \left| \begin{array}{cccc} A_{ii} & A_{ij} & A_{ik} & A_{i\ell} \\
                                  A_{ji} & A_{jj} & A_{jk} & A_{j\ell} \\
                                  A_{ki} & A_{kj} & A_{kk} & A_{k\ell} \\
                                  A_{\ell i} & A_{\ell j} & A_{\ell k} & A_{\ell\ell} 
\end{array} \right| \right| = \nonumber \\
 & & \nonumber \\
 & & \qquad \;\; = \frac{1}{4!} \sum\limits_{i,j,k,\ell=1}^{4} 
(A_{ii} A_{jj} A_{kk} A_{\ell\ell} \pm \ldots ) \nonumber \\
  & & \nonumber \\
 d=5 & \Rightarrow & \det A = \frac{1}{5!} \sum\limits_{i,j,k,\ell ,m=1}^{5} 
\left| \left| \begin{array}{ccccc} A_{ii} & A_{ij} & A_{ik} & A_{i\ell} & A_{im} \\
                                   A_{ji} & A_{jj} & A_{jk} & A_{j\ell} & A_{jm} \\
                                   A_{ki} & A_{kj} & A_{kk} & A_{k\ell} & A_{km} \\
                                   A_{\ell i} & A_{\ell j} & A_{\ell k} & A_{\ell\ell} & A_{\ell m} \\
                                   A_{mi} & A_{mj} & A_{mk} & A_{m\ell} & A_{mm}
\end{array} \right| \right| = \nonumber \\
 & & \nonumber \\
 & & \qquad \;\; = \frac{1}{5!} \sum\limits_{i,j,k,\ell ,m=1}^{5} 
(A_{ii} A_{jj} A_{kk} A_{\ell\ell} A_{mm} \pm \ldots ) \nonumber \\
  & & \nonumber \\
 \cdots & & 
\eeqa

From these expansions it is easy to read off the following expressions which evaluate $\det A$ in 
terms of the $\mbox{tr} A$:

\beqa 
 d=1 & \Rightarrow & \det A = \frac{1}{1!} \; \mbox{tr} A \nonumber \\
 & & \nonumber \\
 d=2 & \Rightarrow & \det A = \frac{1}{2!} \; \big\{ (\mbox{tr} A)^2 
                                                    - \mbox{tr} A^2 \big\} \nonumber \\
 & & \nonumber \\
 d=3 & \Rightarrow & \det A = \frac{1}{3!} \; \big\{ (\mbox{tr} A)^3 
                                                    - \mbox{tr} A \; \mbox{tr} A^2 
                                               + 2 \; \mbox{tr} A^3 \big\} \nonumber \\
 & & \nonumber \\
 d=4 & \Rightarrow & \det A = \frac{1}{4!} \; \big\{ (\mbox{tr} A)^4 
                                              - 6 \; (\mbox{tr} A)^2 \; \mbox{tr} A^2 
                                               + 8 \; \mbox{tr} A \; \mbox{tr} A^3 
                                              + 3 \; (\mbox{tr} A^2)^2  
                                               - 6 \; \mbox{tr} A^4 \big\} \nonumber \\
 & & \nonumber \\
 d=5 & \Rightarrow & \det A = \frac{1}{5!} \; \big\{ (\mbox{tr} A)^5 
                                             - 10 \; (\mbox{tr} A)^3 \; \mbox{tr} A^2 
                                             + 20 \; (\mbox{tr} A)^2 \; \mbox{tr} A^3 
                                              - 30 \; \mbox{tr} A \; \mbox{tr} A^4 - \nonumber \\
                    & & \qquad \;\; \qquad \; - 20 \; \mbox{tr} A^2 \; \mbox{tr} A^3 
                                              + 15 \; \mbox{tr} A \; (\mbox{tr} A^2)^2  
                                              + 24 \; \mbox{tr} A^5 \big\} \nonumber \\
 & & \nonumber \\
 \cdots & & \label{ddt1}
\eeqa

For permanents the minus signs in the upper sums just have to be replaced by plus signs.
In general we obtain the following formula for arbitrary dimension $d$:

\beqa \det A & = & \frac{1}{d!} \; \sum\limits_{1 k_1+2 k_2+\ldots + d k_d=d} \; 
 (-1)^{d- k_1-\ldots -k_d} \;\frac{d!}{1^{k_1} \ldots d^{k_d} \, k_1! \ldots k_d!} \;
 {\left( \mbox{tr} A^1 \right)}^{k_1} \ldots {\left( \mbox{tr} A^d \right)}^{k_d} \nonumber \\
 & & \nonumber \\
 & & (k_1, \ldots , k_d \in \{ 0,1,2, \ldots \}) 
\eeqa

For permanents we obviously get:

\beqa \mbox{per} \, A & = & \frac{1}{d!} \; \sum\limits_{1 k_1+2 k_2+\ldots + d k_d=d} \; 
 \frac{d!}{1^{k_1} \ldots d^{k_d} \, k_1! \ldots k_d!} \;
 {\left( \mbox{tr} A^1 \right)}^{k_1} \ldots {\left( \mbox{tr} A^d \right)}^{k_d} \nonumber \\
 & & \nonumber \\
 & & (k_1, \ldots , k_d \in \{ 0,1,2, \ldots \}) 
\eeqa

The coefficients $d!/(1^{k_1} \ldots d^{k_d} \, k_1! \ldots k_d!)$ are discussed e.g.\ in 
\cite[p.\ 44]{lud1}.

%
% ==========================================================================
%
\section{Expansion of $4 \times 4$ matrices in terms of Dirac covariants}

As a complete linear independent set of sixteen $4 \times 4$ matrices we can use the following
so called Dirac covariants ($g^{\mu\nu}$ is the metric tensor):

\beqa 
 1_4 & & \nonumber \\
 \gamma_5 & & \nonumber \\
 \gamma^\mu & & (\mu = 0,1,2,3), \nonumber \\
 i \gamma_5 \,\gamma^\mu & & (\mu = 0,1,2,3) \nonumber \\
 \sigma^{\mu\nu} & & (\mu < \nu) (\mu , \nu = 0,1,2,3)
\eeqa

($\gamma^\mu \gamma^\nu + \gamma^\nu \gamma^\mu := 2 g^{\mu\nu} 1_4$,
$\;\gamma_5:=i\gamma^0\gamma^1\gamma^2\gamma^3$, 
$\;\sigma^{\mu\nu}=-\sigma^{\nu\mu}:= \frac{i}{2} (\gamma^\mu \gamma^\nu - \gamma^\nu \gamma^\mu)$ 
\cite[p.\ 692]{itz1}) \\

Apart from the 4 dimensional unit--matrix $1_4$ all Dirac covariants are traceless.
All Dirac covariants are trace orthogonal, i.e. if $\Gamma$ and $\Gamma^\prime$ are two
Dirac covariants we obtain:

\beq \mbox{tr} (\Gamma \Gamma^\prime ) = 0 \quad \mbox{for} \quad \Gamma \neq \Gamma^\prime \eeq

Very useful are the following traces:

\beqa 
\mbox{tr} (1_4 1_4) & = & 4 \nonumber \\
\mbox{tr} (\gamma_5 \gamma_5) & = & 4 \nonumber \\
\mbox{tr} (\gamma^\mu \gamma^\nu ) & = & 4 \, g^{\mu\nu} \nonumber \\
\mbox{tr} (i\gamma_5 \gamma^\mu \, i \gamma_5 \gamma^\nu ) & = & 4 \, g^{\mu\nu} \nonumber \\
\mbox{tr} (\sigma^{\mu\nu} \sigma^{\rho\sigma} ) & = & 4 \, (g^{\mu\rho} g^{\nu\sigma} - g^{\mu\sigma}
g^{\nu\rho} ) 
\eeqa

Using the notation $\slsh{a}:=a_{\mu} \gamma^{\mu}$ (Einstein sum convention!) and the completeness of the
Dirac covariants we can expand every $4 \times 4$ matrix $M$ in terms of the Dirac covariants:

\beq M = A \, 1_4 + B \gamma_5 + \slsha{C} + i \gamma_5 \slsha{D} + \frac{1}{2} \, 
E_{\mu\nu} \, \sigma^{\mu\nu} \quad (E_{\mu\nu} = - E_{\nu\mu}) \eeq

The coefficients are easily calculated by the trace properties of the Dirac covariants:

\beqa 
 A & = & \frac{1}{4} \; \mbox{tr} (M) \nonumber \\
 & & \nonumber \\
 B & = & \frac{1}{4} \; \mbox{tr} (M \gamma_5) \nonumber \\
 & & \nonumber \\
 C^{\mu} & = & \frac{1}{4} \; \mbox{tr} (M \gamma^{\mu}) \nonumber \\
 & & \nonumber \\
 D^{\mu} & = & \frac{1}{4} \; \mbox{tr} (M i \gamma_5 \gamma^{\mu}) \nonumber \\
 & & \nonumber \\
 E^{\mu\nu} & = & \frac{1}{4} \; \mbox{tr} (M \sigma^{\mu\nu})  
\eeqa

Products, commutators and anticommutators of Dirac covariants evaluated and expanded in terms of 
Dirac covariants are given in appendix \ref{app1}.

%
% ==========================================================================
%
\section{Inversion of $4 \times 4$ matrices in terms of Dirac covariants}

\subsection{General case}

Like in the previous section we consider an arbitrary $4 \times 4$ matrix expanded in terms 
of Dirac covariants:

\beq M = A \, 1_4 + B \gamma_5 + \slsha{C} + i \gamma_5 \slsha{D} + \frac{1}{2} \, 
E_{\mu\nu} \, \sigma^{\mu\nu} \quad (E_{\mu\nu} = - E_{\nu\mu}) \eeq

Now we define the following shorthand notations:

\beqa E^\ast_{\mu\nu} & := & \varepsilon_{\mu\nu\rho\sigma} \, E^{\rho\sigma} \quad (\mbox{dual tensor of}\,
E^{\mu\nu}) \nonumber \\
 & & \nonumber \\
 C \cdot D & := & C_\mu D^\mu \nonumber \\
 & & \nonumber \\
 C^2 & := & C_\mu C^\mu \quad , \quad D^2 \quad := \quad D_\mu D^\mu \nonumber \\
 & & \nonumber \\
 (D \pm i C)^2 & := & (D_\mu \pm i C_\mu) (D^\mu \pm i C^\mu) \nonumber \\
 & & \nonumber \\
 E^2 & := & E_{\mu\nu} E^{\mu\nu} \nonumber \\
 & & \nonumber \\
 E^\ast E & := & E^\ast_{\mu\nu} E^{\mu\nu} \quad =: E E^\ast
\eeqa

and observe the following identities:

\beq (D-iC)^2  (D+iC)^2 \; = \; (D^2 - C^2)^2 + 4 \, (C \cdot D)^2 \; = \; (D^2 + C^2)^2 + 4 \, (C \cdot D)^2 - 4 \, C^2 D^2 \eeq
\beqa 
 {\left( \frac{1}{2} \, \sigma_{\mu\nu} E^{\mu\nu}\right) }^2 & = & \frac{1}{2} \, E^2 +  \frac{i}{4} \, \gamma_5 E E^\ast \\
 & & \nonumber \\
 \left[ \slsha{C} - i \gamma_5 \slsha{D} \right] \, \left[ \slsha{C} - i \gamma_5 \slsha{D} \right] & = & 
 C^2 + D^2 + 2 \gamma_5 \sigma_{\mu\nu} \; C^\mu D^\nu \\
 & & \nonumber \\
 \left[ \slsha{C} - i \gamma_5 \slsha{D} \right] \, \left[ \slsha{C} + i \gamma_5 \slsha{D} \right] & = & 
 C^2 - D^2 - 2 i \, \gamma_5 \; C \cdot D
\eeqa

Using the trace formula (\ref{ddt1}) for $d=4$ you can verify that:

\beqa \lefteqn{\det M \quad =} \nonumber \\
 & & \nonumber \\ 
 & = & (D - i C)^2 (D + i C)^2 
                 + \left[ (A-B)^2 - \frac{1}{2} \, E^2 + \frac{i}{4} E E^\ast \right ] \,  
                   \left[ (A+B)^2 - \frac{1}{2} \, E^2 - \frac{i}{4} E E^\ast \right ] - \nonumber \\  
 & & \nonumber \\ 
 & - & 2 \; \Big\{ (A^2 - B^2 + \frac{1}{2} \, E^2) (D^2 + C^2) 
            + 4 \, \left[ \frac{i}{2} A E^\ast_{\mu\nu} - B E_{\mu\nu} \right] \, C^\mu D^\nu - \nonumber \\
 & & \nonumber \\ 
 & - & \quad 2 \, E^\lambda_{\;\;\,\mu} E_{\lambda\nu} (C^\mu C^\nu + D^\mu D^\nu) -
                  E^{\ast\,\lambda}_{\;\;\;\;\,\mu} E_{\lambda\nu} (C^\mu D^\nu - D^\mu C^\nu) \Big\}
\eeqa 

It is somewhat involved to derive the following expression for the inverse matrix $M^{-1}$:

\beqa 
 M^{-1} = \frac{1}{\det M} & \Bigg\{ &
 \Bigg[ (A + B \gamma_5 - \frac{1}{2} \, \sigma_{\mu\nu} \, E^{\mu\nu}) (A - B \gamma_5) - \nonumber \\
 & & \nonumber \\
 & - & \quad \left[ \slsha{C} - i \gamma_5 \slsha{D} \right] \, \left[ \slsha{C} - i \gamma_5 \slsha{D} \right] \Bigg] \,
 \left[ A - B \gamma_5 - \slsha{C} - i \gamma_5 \slsha{D} \right] - \nonumber \\
 & & \nonumber \\
 & - & (A + B \gamma_5 - \frac{1}{2} \, \sigma_{\mu\nu} \, E^{\mu\nu})  \,
 \Bigg[ \frac{1}{2} \, E^2 - \frac{i}{4} \, \gamma_5 E E^\ast - 
 \left[ \slsha{C} + i \gamma_5 \slsha{D} \right] 
 \, \frac{1}{2} \, \sigma_{\alpha\beta} \, E^{\alpha\beta} \Bigg]  \nonumber \\
 & & \nonumber \\
 & - & 
 \left[ \slsha{C} - i \gamma_5 \slsha{D} \right] \, \frac{1}{2} \, \sigma_{\mu\nu} \, E^{\mu\nu} \,
 \left[ \slsha{C} - i \gamma_5 \slsha{D} \right] \Bigg\} \nonumber \\
\eeqa

For some purposes the following expression is more suitable:

\beqa 
 M^{-1} = \frac{1}{\det M} & \Bigg\{ &
 (A + B \gamma_5 - \frac{1}{2} \, \sigma_{\mu\nu} \, E^{\mu\nu}) 
 \Bigg[ (A - B \gamma_5)^2 - \frac{1}{2} \, E^2 + \frac{i}{4} \, \gamma_5 E E^\ast - \nonumber \\
 & & \nonumber \\
 & - & \left[ \slsha{C} + i \gamma_5 \slsha{D} \right] 
 (A + B \gamma_5 - \frac{1}{2} \, \sigma_{\alpha\beta} \, E^{\alpha\beta}) \Bigg] - \nonumber \\
 & & \nonumber \\
 & - & \left[ \slsha{C} - i \gamma_5 \slsha{D} \right] \, \Bigg[ 
 (A + B \gamma_5 + \frac{1}{2} \, \sigma_{\alpha\beta} \, E^{\alpha\beta}) \, 
 \left[ \slsha{C} - i \gamma_5 \slsha{D} \right] - \nonumber \\
 & & \nonumber \\
 & - & \left[ \slsha{C} - i \gamma_5 \slsha{D} \right] 
 \,\left[ \slsha{C} + i \gamma_5 \slsha{D} \right] \Bigg] \Bigg\} 
 \eeqa

\subsection{The case $E^{\mu\nu}=0$}

In this subsection we take the limit $E^{\mu\nu}=0$, i.e.:

\beq M = A \, 1_4 + B \gamma_5 + \slsha{C} + i \gamma_5 \slsha{D} \eeq

In this limit we obtain for $\det M$:

\beqa \lefteqn{\det M \quad =} \nonumber \\
 & & \nonumber \\
 & = & (A^2 - B^2 - C^2 - D^2)^2 + 4 \, (C \cdot D)^2 - 4 \, C^2 D^2 = \nonumber \\
 & & \nonumber \\
 & = & (A^2 - B^2)^2 - 2 \, (A^2 - B^2)(C^2 + D^2) + (C^2 + D^2)^2  
  + 4 \, (C \cdot D)^2 - 4 \, C^2 D^2 = \nonumber \\
 & & \nonumber \\
 & = & (A^2 - B^2)^2 - 2 \, (A^2 - B^2)(C^2 + D^2) + (C^2 - D^2)^2  
  + 4 \, (C \cdot D)^2 = \nonumber \\
 & & \nonumber \\
 & = & (A^2 - B^2)^2 - 2 \, (A^2 - B^2)(C^2 + D^2) + (C - i D)^2 (C + i D)^2 = \nonumber \\
 & & \nonumber \\
 & = & (A^2 - B^2 - 2 C^2 - 2 C^2)(A^2 - B^2) + (C - i D)^2 (C + i D)^2 
\eeqa

The corresponding inverse matrix $M^{-1}$ is:

\beqa
 M^{-1} & \displaystyle = \frac{1}{\det M} & 
 \Bigg[ A^2 - B^2 - \left[ \slsha{C} - i \gamma_5 \slsha{D} \right] \,
  \left[ \slsha{C} - i \gamma_5 \slsha{D} \right] \Bigg] \,
  \left[ A - B \gamma_5 -\slsha{C} - i \gamma_5 \slsha{D} \right] \nonumber \\ 
 & & \nonumber \\
 & \displaystyle = \frac{1}{\det M} & 
 \Bigg[ A^2 - B^2 - C^2 - D^2 - 2 \gamma_5 \sigma_{\mu\nu} \; C^\mu D^\nu \Bigg] \,
  \left[ A - B \gamma_5 -\slsha{C} - i \gamma_5 \slsha{D} \right] \nonumber \\
  & & 
\eeqa

If $C^\mu$ is proportional to $D^\mu$, i.e.\ $C^\mu \sim D^\mu$, we get significant simplifications.
We note that:

\beq C^\mu \sim D^\mu \quad \Rightarrow \quad 
4 \, (C \cdot D)^2 - 4 \, C^2 D^2 = 0 \quad , \quad \sigma_{\mu\nu} \; C^\mu D^\nu = 0 \eeq

i.e.

\beq C^\mu \sim D^\mu \quad \Rightarrow \quad \det M = (A^2 - B^2 - C^2 - D^2)^2
\eeq 

For the inverse matrix we obtain:

\beqa C^\mu \sim D^\mu & \Rightarrow \quad M^{-1} & = 
\frac{1}{\det M} (A^2 - B^2 - C^2 - D^2) \,\left[ A - B \gamma_5 -\slsha{C} - i \gamma_5 \slsha{D} \right] = 
\nonumber \\
 & & \nonumber \\
 & & = \frac{A - B \gamma_5 -\slsha{C} - i \gamma_5 \slsha{D}}{A^2 - B^2 - C^2 - D^2}
\eeqa

Some special examples with respect to this case are:

\beqa 
(A + B \gamma_5 + \slsha{C})^{-1} & = & 
\frac{A - B \gamma_5 -\slsha{C}}{A^2 - B^2 - C^2} \nonumber \\
 & & \nonumber \\
(A + B \gamma_5 + i \gamma_5 \slsha{D})^{-1} & = & 
\frac{A - B \gamma_5 - i \gamma_5 \slsha{D}}{A^2 - B^2 - D^2}
\eeqa

Another interesting situation is the case $A=\pm \,B$, i.e.\ we start from the matrix:

\beq M = A \, ( 1_4 \pm \gamma_5) + \slsha{C} + i \gamma_5 \slsha{D} \eeq

Then we get for $\det M$:

\beq \det M \quad = \quad (C - i D)^2 (C + i D)^2 \eeq

The inverse $M^{-1}$ is given by:

\beqa
M^{-1} = & \displaystyle \frac{- 1}{\det M} & 
  \left[ \slsha{C} - i \gamma_5 \slsha{D} \right] \,
  \left[ \slsha{C} - i \gamma_5 \slsha{D} \right] \,
  \left[ A \, ( 1_4 \mp \gamma_5) - \slsha{C} - i \gamma_5 \slsha{D} \right] 
\eeqa

\subsection{The case $C^{\mu}=D^{\mu}=0$}

In this subsection we take the limit $C^{\mu}=D^{\mu}=0$, i.e.:

\beq M = A \, 1_4 + B \gamma_5 + \frac{1}{2} \, E_{\mu\nu} \, \sigma^{\mu\nu} \quad (E_{\mu\nu} = - E_{\nu\mu}) \eeq

In this limit we obtain for $\det M$:

\beqa \det M & = & 
   \left[ (A-B)^2 - \frac{1}{2} \, E^2 + \frac{i}{4} E E^\ast \right ] \,  
   \left[ (A+B)^2 - \frac{1}{2} \, E^2 - \frac{i}{4} E E^\ast \right ]    
\eeqa 

The inverse matrix $M^{-1}$ is given by:

\beqa 
 M^{-1} & = & \frac{1}{\det M} \,
 (A + B \gamma_5 - \frac{1}{2} \, \sigma_{\mu\nu} \, E^{\mu\nu}) \,
 \Bigg[ (A - B \gamma_5)^2 - \frac{1}{2} \, E^2 + \frac{i}{4} \, \gamma_5 E E^\ast  \Bigg] = \nonumber \\ 
 & & \nonumber \\
 & = & \frac{1}{\det M} \,
 \Bigg[ (A - B \gamma_5)^2 - \frac{1}{2} \, E^2 + \frac{i}{4} \, \gamma_5 E E^\ast  \Bigg] \, 
 (A + B \gamma_5 - \frac{1}{2} \, \sigma_{\mu\nu} \, E^{\mu\nu}) 
\eeqa

%
% Anhang:
% =======                              
%

\begin{appendix}
\section{Products, commutators and anticommutators of Dirac covariants} \label{app1}

The totally antisymmetric Levi-Civita tensor is defined by \cite[p.\ 692]{itz1}:

\beq \varepsilon_{\mu\nu\rho\sigma} := 
\left| \left| \begin{array}{cccc} g_{\mu 0} & g_{\mu 1} & g_{\mu 2} & g_{\mu 3} \\
                                  g_{\nu 0} & g_{\nu 1} & g_{\nu 2} & g_{\nu 3} \\
                                  g_{\rho 0} & g_{\rho 1} & g_{\rho 2} & g_{\rho 3} \\
                                  g_{\sigma 0} & g_{\sigma 1} & g_{\sigma 2} & g_{\sigma 3} 
\end{array} \right| \right| \eeq

For the commutator and anticommutator of $A$ and $B$ we write:

\beq [A,B] := AB-BA \quad , \quad \{ A,B \}:= AB+BA \eeq

Some calculation yields the following expansions of products of Dirac covariants in terms
of Dirac covariants ($\varepsilon^{\mu\nu}_{\quad \lambda\lambda^\prime} \; \varepsilon^{\lambda\lambda^\prime\rho\sigma}
\stackrel{!}{=} -2\;(g^{\mu\rho} g^{\nu\sigma}- g^{\mu\sigma} g^{\nu\rho})$):

\[ \begin{array}{|ccl|ccl|ccl|} \hline
 & & & & & & & & \\
 1_4 \cdot 1_4 & = & 1_4 & 
 1_4 \cdot \gamma_5 & = & \gamma_5 & 
 1_4 \cdot \gamma_\mu & = & \gamma_\mu \\
 & & & & & & & & \\ \hline
 & & & & & & & & \\
 \gamma_5 \cdot 1_4 & = & \gamma_5 & 
 \gamma_5 \cdot \gamma_5 & = & 1_4 & 
 \gamma_5 \cdot \gamma_\mu & = & -i (i \gamma_5 \gamma_\mu) \\ 
 & & & & & & & & \\ \hline
 & & & & & & & & \\
 \gamma_\mu \cdot 1_4 & = & \gamma_\mu & 
 \gamma_\mu \cdot \gamma_5 & = & i (i \gamma_5 \gamma_\mu) & 
 \gamma_\mu \cdot \gamma_\nu & = &  g_{\mu\nu} 1_4 - i \sigma_{\mu\nu} \\ 
 & & & & & & & & \\ \hline
 & & & & & & & & \\
 i \gamma_5 \gamma_\mu \cdot 1_4 & = & i \gamma_5 \gamma_\mu & 
 i \gamma_5 \gamma_\mu \cdot \gamma_5 & = & - i \gamma_\mu & 
 i \gamma_5 \gamma_\mu \cdot \gamma_\nu & = &  i \; g_{\mu\nu} \gamma_5 + \frac{i}{2} \, \varepsilon_{\mu\nu\rho\sigma} \; \sigma^{\rho\sigma} \\ 
 & & & & & & & & \\ \hline
 & & & & & & & & \\
 \sigma_{\mu\nu} \cdot 1_4 & = & \sigma_{\mu\nu} & 
 \sigma_{\mu\nu} \cdot \gamma_5 & = & \frac{i}{2} \, \varepsilon_{\mu\nu\rho\sigma} \; \sigma^{\rho\sigma} & 
 \sigma_{\mu\nu} \cdot \gamma_\rho & = & i \varepsilon_{\sigma\mu\nu\rho} \; i \gamma_5 \gamma^{\sigma} + \\
 & & & & & & & & \\ 
 & & &
 & & &
 & + & \frac{i}{2} \, \varepsilon_{\mu\nu\lambda\lambda^\prime} \; \varepsilon^{\lambda\lambda^\prime}_{\quad\rho\sigma}
     \; \gamma^\sigma \\  
 & & & & & & & & \\ \hline
\end{array}
\]

\[ \begin{array}{|ccl|ccl|} \hline
 & & & & & \\
 1_4 \cdot i \gamma_5 \gamma_{\mu} & = &  i \gamma_5 \gamma_{\mu} & 
 1_4 \cdot \sigma_{\mu\nu} & = &  \sigma_{\mu\nu} \\ 
 & & & & & \\ \hline
 & & & & & \\
 \gamma_5 \cdot i \gamma_5 \gamma_{\mu} & = &  i \gamma_{\mu} & 
 \gamma_5 \cdot \sigma_{\mu\nu} & = & \frac{i}{2} \, \varepsilon_{\mu\nu\rho\sigma} \; \sigma^{\rho\sigma} \\
 & & & & & \\ \hline
 & & & & & \\
 \gamma_\mu \cdot i \gamma_5 \gamma_{\nu} & = & - i g_{\mu\nu} \gamma_5 - \frac{i}{2} \, \varepsilon_{\mu\nu\rho\sigma} \; \sigma^{\rho\sigma} & 
 \gamma_\rho \cdot \sigma_{\mu\nu} & = & i \, \varepsilon_{\sigma\mu\nu\rho} \; i \gamma_5 \gamma^{\sigma} - \\ 
 & & & & & \\ 
 & & &
 & - & \frac{i}{2} \, \varepsilon_{\mu\nu\lambda\lambda^\prime} \; \varepsilon^{\lambda\lambda^\prime}_{\quad\rho\sigma}
     \; \gamma^\sigma \\  
 & & & & & \\ \hline
 & & & & & \\
 i \gamma_5 \gamma_\mu \cdot i \gamma_5 \gamma_\nu & = &  g_{\mu\nu} 1_4 - i \sigma_{\mu\nu} & 
 i \gamma_5 \gamma_\rho \cdot \sigma_{\mu\nu} & = & - i \varepsilon_{\sigma\mu\nu\rho} \; \gamma^{\sigma} - \\
 & & & & & \\ 
 & & &
 & - & \frac{i}{2} \, \varepsilon_{\mu\nu\lambda\lambda^\prime} \; \varepsilon^{\lambda\lambda^\prime}_{\quad\rho\sigma}
     \; i \gamma_5 \gamma^\sigma \\  
 & & & & & \\ \hline
 & & & & & \\
 \sigma_{\mu\nu} \cdot i \gamma_5 \gamma_\rho & = & - i \varepsilon_{\sigma\mu\nu\rho} \; \gamma^{\sigma} + &
 \sigma_{\mu\nu} \cdot \sigma_{\rho\sigma} & = & (g_{\mu\rho} g_{\nu\sigma}- g_{\mu\sigma} g_{\nu\rho}) \; 1_4 + \\
  & & & & & \\ 
 & + & \frac{i}{2} \, \varepsilon_{\mu\nu\lambda\lambda^\prime} \; \varepsilon^{\lambda\lambda^\prime}_{\quad\rho\sigma}
     \; i \gamma_5 \gamma^\sigma &  
 & + & i \; \left( \sigma_{\mu\sigma} g_{\nu\rho}- \sigma_{\mu\rho} g_{\nu\sigma} \right. \; + \\
 & & & & & \\
 & & &
 & + & \quad \left. g_{\mu\sigma} \sigma_{\nu\rho} - g_{\mu\rho} \sigma_{\nu\sigma} \right) + \\
 & & & & & \\
 & & &
 & + & i \; \varepsilon_{\mu\nu\rho\sigma} \; \gamma_5 \\
 & & & & & \\ \hline
\end{array}
\]
\beq \eeq

The commutators of the Dirac covariants are evaluated to be:

\[ \begin{array}{|ccl|ccl|ccl|} \hline
 & & & & & & & & \\
 {[ 1_4 , 1_4 ]} & = & 0 & 
 {[ 1_4 , \gamma_5 ]} & = & 0 & 
 {[ 1_4 , \gamma_\mu ]} & = & 0 \\
 & & & & & & & & \\ \hline
 & & & & & & & & \\
 {[ \gamma_5 , 1_4 ]} & = & 0 & 
 {[ \gamma_5 , \gamma_5 ]} & = & 0 & 
 {[ \gamma_5 , \gamma_\mu ]} & = & - 2 i (i \gamma_5 \gamma_\mu) \\ 
 & & & & & & & & \\ \hline
 & & & & & & & & \\
 {[ \gamma_\mu , 1_4 ]} & = & 0 & 
 {[ \gamma_\mu , \gamma_5 ]} & = & 2 i (i \gamma_5 \gamma_\mu) & 
 {[ \gamma_\mu , \gamma_\nu ]} & = &  - 2 i \sigma_{\mu\nu} \\ 
 & & & & & & & & \\ \hline
 & & & & & & & & \\
 {[ i \gamma_5 \gamma_\mu , 1_4 ]} & = & 0 & 
 {[ i \gamma_5 \gamma_\mu , \gamma_5 ]} & = & - 2 i \gamma_\mu & 
 {[ i \gamma_5 \gamma_\mu , \gamma_\nu ]} & = &  2 i \; g_{\mu\nu} \gamma_5 \\ 
 & & & & & & & & \\ \hline
 & & & & & & & & \\
 {[ \sigma_{\mu\nu} , 1_4 ]} & = & 0 & 
 {[ \sigma_{\mu\nu} , \gamma_5 ]} & = & 0 & 
 {[ \sigma_{\rho\sigma} , \gamma_\mu ]} & = & 2 i \; ( \gamma_\rho g_{\mu\sigma} - \gamma_\sigma g_{\mu\rho} ) \\
 & & & & & & & & \\ \hline
\end{array}
\]

\[ \begin{array}{|ccl|ccl|} \hline
 & & & & & \\
 {[ 1_4 , i \gamma_5 \gamma_{\mu} ]} & = & 0 & 
 {[ 1_4 , \sigma_{\mu\nu} ]} & = & 0 \\ 
 & & & & & \\ \hline
 & & & & & \\
 {[ \gamma_5 , i \gamma_5 \gamma_{\mu} ]} & = &  2 i \gamma_{\mu} & 
 {[ \gamma_5 , \sigma_{\mu\nu} ]} & = & 0 \\
 & & & & & \\ \hline
 & & & & & \\
 {[ \gamma_\mu , i \gamma_5 \gamma_{\nu} ]} & = & - 2 i g_{\mu\nu} \gamma_5 & 
 {[ \gamma_\mu , \sigma_{\rho\sigma} ]} & = & 2 i \; ( \gamma_\sigma g_{\mu\rho} - \gamma_\rho g_{\mu\sigma} )  \\ 
 & & & & & \\ \hline
 & & & & & \\
 {[ i \gamma_5 \gamma_\mu , i \gamma_5 \gamma_\nu ]} & = &  - 2 i \sigma_{\mu\nu} & 
 {[ i \gamma_5 \gamma_\mu , \sigma_{\rho\sigma} ]} & = & 2 i \; ( i \gamma_5 \gamma_\sigma g_{\mu\rho} - i \gamma_5 \gamma_\rho g_{\mu\sigma} )  \\
 & & & & & \\ \hline
 & & & & & \\
 {[ \sigma_{\rho\sigma} , i \gamma_5 \gamma_\mu ]} & = & 2 i \; ( i \gamma_5 \gamma_\rho g_{\mu\sigma} - i \gamma_5 \gamma_\sigma g_{\mu\rho} )  &
 {[ \sigma_{\mu\nu} , \sigma_{\rho\sigma} ]} & = &  2 i \; \left( \sigma_{\mu\sigma} g_{\nu\rho}- \sigma_{\mu\rho} g_{\nu\sigma} \right. \; + \\
  & & & & & \\ 
 & & &
 & + & \quad \; \left. g_{\mu\sigma} \sigma_{\nu\rho} - g_{\mu\rho} \sigma_{\nu\sigma} \right) \\
 & & & & & \\ \hline
\end{array}
\]
\beq \eeq

For the anticommutators of the Dirac covariants we obtain:

\[ \begin{array}{|ccl|ccl|ccl|} \hline
 & & & & & & & & \\
 \{ 1_4 , 1_4 \} & = & 2 \, 1_4 & 
 \{ 1_4 , \gamma_5 \} & = & 2 \, \gamma_5 & 
 \{ 1_4 , \gamma_\mu \} & = & 2 \, \gamma_\mu \\
 & & & & & & & & \\ \hline
 & & & & & & & & \\
 \{ \gamma_5 , 1_4 \} & = & 2 \, \gamma_5 & 
 \{ \gamma_5 , \gamma_5 \} & = & 2 \, 1_4 & 
 \{ \gamma_5 , \gamma_\mu \} & = & 0 \\ 
 & & & & & & & & \\ \hline
 & & & & & & & & \\
 \{ \gamma_\mu , 1_4 \} & = & 2 \, \gamma_\mu & 
 \{ \gamma_\mu , \gamma_5 \} & = & 0 & 
 \{ \gamma_\mu , \gamma_\nu \} & = &  2 \, g_{\mu\nu} 1_4 \\ 
 & & & & & & & & \\ \hline
 & & & & & & & & \\
 \{ i \gamma_5 \gamma_\mu , 1_4 \} & = & 2  i \gamma_5 \gamma_\mu & 
 \{ i \gamma_5 \gamma_\mu , \gamma_5 \} & = & 0 & 
 \{ i \gamma_5 \gamma_\mu , \gamma_\nu \} & = &  i \, \varepsilon_{\mu\nu\rho\sigma} \; \sigma^{\rho\sigma} \\
 & & & & & & & & \\ \hline
 & & & & & & & & \\
 \{ \sigma_{\mu\nu} , 1_4 \} & = & 2 \, \sigma_{\mu\nu} & 
 \{ \sigma_{\mu\nu} , \gamma_5 \} & = & i \, \varepsilon_{\mu\nu\rho\sigma} \; \sigma^{\rho\sigma} & 
 \{ \sigma_{\mu\nu} , \gamma_\rho \} & = & 2 i \varepsilon_{\sigma\mu\nu\rho} \; i \gamma_5 \gamma^{\sigma} \\
 & & & & & & & & \\ \hline
\end{array}
\]

\[ \begin{array}{|ccl|ccl|} \hline
 & & & & & \\
 \{ 1_4 , i \gamma_5 \gamma_{\mu} \} & = &  2 i \, \gamma_5 \gamma_{\mu} & 
 \{ 1_4 , \sigma_{\mu\nu} \} & = &  2 \, \sigma_{\mu\nu} \\ 
 & & & & & \\ \hline
 & & & & & \\
 \{ \gamma_5 , i \gamma_5 \gamma_{\nu} \} & = & 0 & 
 \{ \gamma_5 , \sigma_{\mu\nu} \} & = & i \, \varepsilon_{\mu\nu\rho\sigma} \; \sigma^{\rho\sigma} \\
 & & & & & \\ \hline
 & & & & & \\
 \{ \gamma_\mu , i \gamma_5 \gamma_{\nu} \} & = & - i \, \varepsilon_{\mu\nu\rho\sigma} \; \sigma^{\rho\sigma} & 
 \{ \gamma_\rho , \sigma_{\mu\nu} \} & = & 2 i \, \varepsilon_{\sigma\mu\nu\rho} \; i \gamma_5 \gamma^{\sigma} \\ 
 & & & & & \\ \hline
 & & & & & \\
 \{ i \gamma_5 \gamma_\mu , i \gamma_5 \gamma_\nu \} & = &  2 g_{\mu\nu} 1_4 & 
 \{ i \gamma_5 \gamma_\rho , \sigma_{\mu\nu} \} & = & - 2 i \, \varepsilon_{\sigma\mu\nu\rho} \; \gamma^{\sigma} \\
 & & & & & \\ \hline
 & & & & & \\
 \{ \sigma_{\mu\nu} , i \gamma_5 \gamma_\rho \} & = & - 2 i \, \varepsilon_{\sigma\mu\nu\rho} \; \gamma^{\sigma} &
 \{ \sigma_{\mu\nu} , \sigma_{\rho\sigma} \} & = & 2 \; (g_{\mu\rho} g_{\nu\sigma}- g_{\mu\sigma} g_{\nu\rho}) \; 1_4 + 
 2 i \; \varepsilon_{\mu\nu\rho\sigma} \; \gamma_5 \\
 & & & & & \\ \hline
\end{array}
\]
\beq \eeq

\end{appendix}
%
%
% Literaturverzeichnis:
% =====================
%

%

\end{document}